\documentclass[sigconf]{acmart}

\AtBeginDocument{%
  \providecommand\BibTeX{{%
    \normalfont B\kern-0.5em{\scshape i\kern-0.25em b}\kern-0.8em\TeX}}}

\copyrightyear{2022}
\acmYear{2022}
\setcopyright{acmcopyright}\acmConference[ASSETS '22]{The 24th International ACM SIGACCESS Conference on Computers and Accessibility}{October 23--26, 2022}{Athens, Greece}
\acmBooktitle{The 24th International ACM SIGACCESS Conference on Computers and Accessibility (ASSETS '22), October 23--26, 2022, Athens, Greece}
\acmPrice{15.00}
\acmDOI{10.1145/3517428.3544821}
\acmISBN{978-1-4503-9258-7/22/10}

\newcommand*{\upnotes}{\includegraphics[scale=0.4]{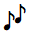}}%

\newcommand*{\downnotes}{\includegraphics[scale=0.4]{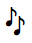}}%

\newcommand*{\upsocial}{\includegraphics[scale=0.4]{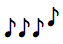}}%

\newcommand*{\downsocial}{\includegraphics[scale=0.4]{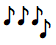}}%

\begin{document}

\title{VRBubble: Enhancing Peripheral Awareness of Avatars for People with Visual Impairments in Social Virtual Reality}

\author{Tiger Ji}
\email{tfji@wisc.edu}
\affiliation{
    \institution{University of Wisconsin-Madison}
    \city{Madison}
    \state{Wisconsin}
    \country{USA}
}

\author{Brianna R. Cochran}
\email{bcochran2@wisc.edu}
\affiliation{
    \institution{University of Wisconsin-Madison}
    \city{Madison}
    \state{Wisconsin}
    \country{USA}
}

\author{Yuhang Zhao}
\email{yuhang.zhao@cs.wisc.edu}
\affiliation{
    \institution{University of Wisconsin-Madison}
    \city{Madison}
    \state{Wisconsin}
    \country{USA}
}

\renewcommand{\shortauthors}{T. Ji et al.}
\renewcommand{\shorttitle}{VRBubble}

\newcommand{\TODO}[1]{\textcolor{red}{\textbf{[#1]}}}

\begin{abstract}
  Social Virtual Reality (VR) is growing for remote socialization and collaboration. However, current social VR applications are not accessible to people with visual impairments (PVI) due to their focus on visual experiences. We aim to facilitate social VR accessibility by enhancing PVI's peripheral awareness of surrounding avatar dynamics. We designed \textit{VRBubble}, an audio-based VR technique that provides surrounding avatar information based on social distances. Based on Hall’s proxemic theory, VRBubble divides the social space with three Bubbles---Intimate, Conversation, and Social Bubble---generating spatial audio feedback to distinguish avatars in different bubbles and provide suitable avatar information.
  We provide three audio alternatives: earcons, verbal notifications, and real-world sound effects. PVI can select and combine their preferred feedback alternatives for different avatars, bubbles, and social contexts. We evaluated VRBubble and an audio beacon baseline with 12 PVI in a navigation and a conversation context. We found that VRBubble significantly enhanced participants' avatar awareness during navigation and enabled avatar identification in both contexts. However, VRBubble was shown to be more distracting in crowded environments.  
\end{abstract}

%




\maketitle

\section{Introduction}
Social virtual reality (VR) refers to VR platforms that allow users to socialize with each other in the form of avatars in a virtual space \cite{socialvr}.  
More and more social VR platforms have been deployed to the market, such as VRChat, Rec Room, and Altspace \cite{vrchat, recroom, altspace}.
Compared to 2D video conferencing systems that cause ``Zoom fatigue'' \cite{fatigue}, social VR offers an immersive and engaging experience that can enhance interpersonal interaction and boost productivity. 
As a result, social VR has attracted increasing attention in recent years. The VR market was valued at \$7.81 billion as of 2020, and is expected to grow 28.2\% annually from 2021 to 2028 \cite{VRMarketSize}. Meta, formerly known as Facebook, also re-branded in 2021 alongside their full commitment to producing a Metaverse, envisioning social VR to be the next generation of Internet that connects everyone. 

Unfortunately, similar to most mainstream VR applications, current social VR mainly targets sighted users by providing various visual avatar designs and supporting non-verbal social interactions, which poses barriers to people with visual impairments (PVI). With more than two billion people experiencing visual impairments worldwide \cite{whopvi}, it is vital to provide PVI equal access to the emerging social VR as virtual collaboration and gathering increases, especially during the COVID-19 pandemic \cite{growth}. 

Researchers have started tackling the VR accessibility problems for PVI by enabling them to navigate and perceive VR scenes. Some research leveraged or created additional devices (e.g., PHANToM, game controller thumbsticks) to enable VR navigation by providing haptic feedback and/or audio feedback \cite{phantom, zhao2018enabling, siu2020virtual, locomotion, rendering, colwellhaptic, nair2021navstick}. 
Others focused on software solutions, designing accessible interactions based on existing VR setups, such as keyboard-based interactions with spatial audio feedback for a virtual world \cite{radegast, textsl, trewin2008powerup, swaminathan}, and more intuitive interactions based on off-the-shelf VR controllers and headsets \cite{zhao2019seeingvr}. 
However, prior work mainly focused on basic VR tasks such as navigation and object perception. It does not address the unique barriers caused by the dynamic and multiplayer nature of social VR. 

Different from a static VR scene with only system-generated objects, social VR is more complex and challenging---human controlled avatars constantly move in the environment and interact with other avatars and objects. All the avatar dynamics are mainly presented visually in social VR and not accessible to PVI. To our knowledge, no existing techniques have focused on the avatar dynamics to support accessible social VR experience for PVI. 

We aim to fill this gap by enhancing PVI's awareness of surrounding avatars in social VR. Unlike prior work that required PVI to actively explore and query information from the environment \cite{nair2021navstick, zhao2018enabling}, we focus on \textit{peripheral awareness}---the innate ability to unconsciously ``maintain and constantly update a sense of our social and physical context'' \cite{aroma}. People can usually maintain peripheral awareness effortlessly without being distracted from their main focus \cite{sideshow}. This ability is especially important in social and collaborative environments since it provides more context for one's activity. For example, when navigating to a specific location, a sighted person can easily notice who has passed by to decide whether to greet that person or start a quick conversation; when in the middle of a conversation, they can stay aware of who just joined the conversation or who is close enough to overhear the conversation.  

We seek to facilitate the peripheral awareness of avatars for PVI in social VR. Via an iterative design with six PVI, we designed \textit{VRBubble}, an audio-based VR technique that provides surrounding avatar information based on social distances. Following Hall's proxemic theory \cite{hall}, we split the virtual space with three ``bubbles'' centered on the user---the Intimate Bubble, Conversation Bubble, and Social Bubble---to represent different social spaces that are suitable for different social interactions. VRBubble then provides three spatial audio alternatives (i.e., earcons, real-world sound effects, and verbal notifications) to convey the avatar information, such as names, relationships with the PVI (friends or strangers), and their interactions with the bubbles (entering or leaving). PVI can also flexibly select and combine their preferred audio alternatives for different bubbles and avatars to maintain awareness of the avatar dynamics. 

We evaluated VRBubble with 12 participants with visual impairments in the context of a navigation task and a conversation task in social VR, with an audio beacon attached to each avatar as the baseline. Our study showed that VRBubble enhanced user avatar awareness while navigating and was effective at providing users with previously inaccessible identifying information about avatars. We also found that PVI generally favor receiving verbal descriptions while navigating, and more brief and intuitive sounds while conversing.



\section{Related Work}

\subsection{Accessibility of Virtual Environments}

\subsubsection{Audio Techniques for VR accessibility}
There has been extensive work that assisted PVI in exploring virtual environments via audio feedback. 
We summarize different types of audio techniques in prior work below.

\textit{\textbf{Audio beacons.}} Audio beacons have been used to convey object positions \cite{walker2006navigation,lokki2005navigation, maidenbaum2013increasing, BlindSwordsman}. For example, Walker and Lindsay's \cite{walker2006navigation} study utilized three different audio beacons in navigation guidance. They observed the impacts on PVI's navigation performance as they changed various parameters, such as timbre and distance to a waypoint to trigger the audio. Maidenbaum et al. \cite{maidenbaum2013increasing} provided a beeping sound based on the distance between the PVI's avatar and the virtual object in front of them to facilitate navigation in a virtual space. As the avatar got closer to the object, the beeping rose in frequency of beeps. Blind Swordsman \cite{BlindSwordsman} was a VR game on mobile devices, where a blind user can hear the spatial audio beacon from the enemies, physically turn to that direction, and tap the touchscreen to swing his sword in the direction he is facing. 

\textit{\textbf{Object Sonification.}}
Prior work has also used audio to identify objects within the virtual space \cite{heuten2006sonification, oliveira2015games, sanchez1997hyperstories, sanchez1999virtual, ShadeofDoom, AHeroCall, merabet2009audio, sanchez2006assisting, atkinson2006making}. For example, de Oliveira et al. \cite{oliveira2015games} recreated a virtual stage and placed instruments in the environment that generated spacial music. Participants then listened to the music tracks to identify and locate instruments on the stage. Heuten et al. \cite{heuten2006sonification} presented a sonification interface to virtual maps for PVI. PVI can listen to the spatialized earcons specific to various geographic objects and landmarks on the map to construct a mental model of a geographical area. AudioDoom \cite{sanchez1999virtual, sanchez1997hyperstories} was an acoustic virtual environment designed for blind children. Virtual objects and events generated spatial sounds to help users identify objects, navigate the space, and improve cognitive skills.

\textit{\textbf{Echolocation.}}
Virtual echolocation has been emulated through signals and audio reflections \cite{waters2001bat, wilkerson2010maze, andrade2018echo}. For example, Andrade et al. \cite{andrade2018echo} enabled PVI to use echolocation to navigate a desktop-based virtual world, where the user's avatar can produce mouth-click or clap sounds by pressing a key on a keyboard and hear the sound reflected in the environment. Waters and Abulula \cite{waters2001bat} presented a sonar system based on the reflectivity of ultrasound used in the echolocation of bats. The audio used was scaled within human hearing ranges, so that PVI can utilize the audio to navigate a VR environment. However, the echolocation method was only used by a small amount of blind people.

\textit{\textbf{User Queried Verbal Descriptions}}
Some works allowed users to select objects in the environment via a list or a grid and hear additional descriptions about the selected object \cite{westin2004game, matsuo2016audible, nair2021navstick, trewin2008powerup}. For instance, Terraformers \cite{westin2004game} was a virtual world game designed to be accessible to PVI. It provided a menu for nearby objects. A user can thus navigate the menu and hear audio descriptions of selected objects. Nair et al. \cite{nair2021navstick} also compared similar menu systems with their novel game controller-based navigation technique, allowing PVI to look around a virtual world by scrubbing the thumbstick on a game controller to different directions; the system then announced what was in that direction via spatial verbal descriptions.

\subsubsection{Haptic Solutions for VR accessibility.}
Prior work has also enhanced the accessibility of virtual environments for PVI by generating haptic feedback or creating haptic controllers \cite{jansson1999haptic, tzovaras2002design, tzovaras2009interactive, zhao2018enabling, wedoff2019virtual,siu2020virtual, lahav2004exploration, schloerb2010blindaid, sato2019assist}. For example, Jansson et al. \cite{jansson1999haptic} enabled PVI to use the stylus on a Phantom Premium device to ``touch'' a virtual space and receive force feedback to perceive different virtual surfaces and objects. Tzovaras et al. \cite{tzovaras2002design} leveraged the CyberGrasp haptic gloves to generate force feedback to a blind user’s fingers, providing the illusion that they were navigating the virtual space with a cane. 
In the same vein, Zhao et al. created Canetroller \cite{zhao2018enabling}, a wearable haptic VR controller that simulated white cane interaction for blind people in virtual reality. When a user swept the controller and hit a virtual object, a physical resistance and spatial audio sound effect were generated to simulate the feedback of a white cane hitting a real object. A follow-up work by Siu et al. \cite{siu2020virtual} further improved the design of the controller by providing three dimensional force feedback. Recently, Wedoff et al. designed a virtual reality game, called Virtual Showdown \cite{wedoff2019virtual}. PVI played the game by hitting a virtual ball into the opponent' goal across a virtual table using a bat. In this game, a Kinect was used to track the user's movement. A visually impaired user can then hold a Nintendo Switch controller as the bat and receive both audio and vibration feedback to perceive the relative position of the ball from his bat.

Prior research on VR accessibility focuses on space navigation and object perception. However, social VR introduces additional complications with the dynamic and non-uniform avatars that present social implications. No research has addressed the accessibility of avatars in social VR. Our research aims to fill this gap by facilitating PVI's awareness of avatars in  social VR via customizable audio techniques.

\subsection{Accessibility of the Real World}
As with virtual environments, a myriad of prior work has designed audio techniques to sonify real world environments, including using audio beacons to mark waypoints \cite{wilson2007swan, holland2002gps}, informing users about nearby objects and landmarks through verbal descriptions \cite{sato2019assist, gleason2018footnote}, generating auditory icons or earcons to identify points of interest \cite{may2020spotlights, presti2019watchout}, and providing echolocation or sonar systems to enable the exploration of surrounding environments \cite{vera2014cane, fakhrhosseini2014listen}. Similar to audio techniques for virtual worlds, these solutions also do not address the dynamic complexity of avatars in a social VR context. 

\subsubsection{Technologies to Facilitate Real-World Social Activities}
More relevant research has been done to assist PVI in real-world social activities by enhancing their awareness of the conversational partners and their non-verbal behaviors. Various wearable or handheld assistive technologies have been developed. These technologies came in a variety of forms, including smartphone applications \cite{zhao2018face}, belts\cite{mcdaniel2008using, buimer2018conveying}, gloves \cite{krishna2010vibroglove, krishna2008systematic}, headband \cite{qiu2016designing, morrison2021peoplelens}, glasses \cite{anam2014expression, terven2014robust}, and wearable cameras \cite{gade2009person, terven2014robust}. Some research focused on haptic feedback. For example, Krishna et al. \cite{krishna2008systematic} created a haptic glove called VibroGlove that allowed PVI to understand facial expressions of a conversation partner. The glove consisted of several vibration motors mounted on the back of each finger that was used to present six different emotions (i.e., anger, disgust, happiness, fear, surprise, and sadness) through vibration patterns. These patterns were designed based around the shape of the mouth and eye area of each facial expression. Another example was Tactile Ban, a wearable headband prototyped by Qiu et al. \cite{qiu2016designing}, which used tactile feedback to provide the feeling of other people's gazes onto the PVI. The band had two vibration patterns based on if a person glanced at the PVI or if their gaze was fixated on them currently. 

Some work focused on audio feedback. For example, Anam et al. \cite{anam2014expression} used Google glasses to track faces and communicate social signals (i.e., facial features, behavioral expressions, body postures, and emotions) of surrounding people to the PVI verbally. Recently, Morrison et al. \cite{morrison2021peoplelens} designed PeopleLens, a head-mounted device that provided spatial identifying audio to assist blind children with gaze direction and mental mapping of surrounding people. Bump and woodblock sounds were used to guide the user's gaze to center on a face, while names would be read out for identified people in the surroundings as the user's gaze passed over them.

Prior work has focused on enhancing the primary social tasks, conveying information about the conversational partners. Besides the primary task, peripheral awareness (as a secondary task) is also important in social activities to \textit{unconsciously} sense the surrounding dynamics and make \textit{ad hoc} social decisions. However, this ability remains understudied for PVI. 

\subsubsection{Assistive Peripheral Awareness Technologies}
There has been some prior work that facilitated peripheral awareness during collaborative tasks through visual \cite{sideshow, sakashita2022remote} or audio \cite{jung2008sound, kilander2002ambient, das2022, lee2022collab} cues. For example, Cadiz et al. \cite{sideshow} created the Sideshow interface to support the user's peripheral awareness of information on their computer. The interface remained on the user's primary display and presented personalized information through visual notifications and summaries, such as number of unread emails or number of friends online. Sakashita et al. \cite{sakashita2022remote} enabled remote collaboration involving physical artifacts through the use of video conferencing and motion tracking. They placed video devices to represent each collaborator and automatically oriented devices to emulate the gaze direction of the collaborator. This supported the user's peripheral awareness of what part of the physical artifact a collaborator was focusing on. However, these works focus on sighted people by providing visual feedback. 

Some work designed audio feedback to enhance PVI's peripheral awareness in a work collaboration scenarios \cite{lee2022collab, das2022, jung2008sound}. For example, Lee et al. \cite{lee2022collab} designed the CollabAlly browser extension, which provided blind users with audio feedback to support the navigation of content or comment changes. Earcons were utilized to convey which part of the document was being edited and different voices were used for text-to-speech to contextualize which collaborator was editing. Jung \cite{jung2008sound} used ambient music to convey notifications to individuals within a physical work space. Speakers were placed throughout the space to enable spatial audio, then unique musical notifications were played at an individual's location when events relevant to that individual occurred, such as the reception of an email. 

However, this work does not design for the unique challenges posed by a social context. Compared to collaborative tasks which involve a small number of collaborators and allow for asynchronous interactions, a social context can involve a large number of moving avatars, generating more peripheral information and distraction. Thus, a different design is needed to adequately facilitate peripheral awareness for the social context. 
\section{Design of VRBubble}
We designed \textit{VRBubble}, an audio-based VR interaction technique to enhance the peripheral awareness of avatars in social VR for PVI. Our design followed the method of user-centered design \cite{userdesign}. We first formulated a set of general design considerations based on prior literature to design an initial prototype. With a formative study with six PVI, we further iterated and improved our design \cite{ji2022vrbubble}. We describe our design process and the final design below.

\subsection{General Design Considerations}
We formulated the following design considerations (C1-C3) based on insights from prior literature \cite{grubert2018office, zhao2018face}.

(C1) \textit{Leverage mainstream platforms.} \label{C1} We focus on desktop VR, where a user can see the virtual environment on the desktop screen, hear the spatial audio feedback via earphones or speakers, and interact with it via keyboard and mouse. While VR headsets are emerging, most people don't own a headset due to its cost \cite{costs}, let alone PVI who cannot benefit from the visual feedback from the headset. Instead, desktop computers are more widely used and many social VR platforms (e.g., VRChat, AltSpace) support desktop access. Moreover, following the VR office concept by Grubert et al. \cite{grubert2018office}, stationary VR in front of a desk that can be controlled by a keyboard can support more comfortable and efficient interaction for longer use and higher productivity in social and collaborative context. 

(C2) \textit{Convey proper information about avatar dynamics.} Prior research has explored PVI's needs for information in real-world social activities and indicated that the top two important information include the identity of surrounding people and their relative location \cite{zhao2018face}. We thus translate such needs to the virtual world and provide the corresponding avatar information to support PVI in the social VR context.

(C3) \textit{Avoid intrusiveness and distraction.} 
Since peripheral awareness is supposed to be effortless \cite{sideshow, matthews2005toolkit}, we seek to minimize the distraction of our design as well as the user's conscious effort. We thus consider short but intuitive audio feedback at suitable timing to convey avatar information to PVI in an unobtrusive manner. Moreover, instead of having users to actively query information, we focus on proactive notifications to reduce their interaction effort. 

\subsection{VRBubble based on Hall's Proxemic Theory} 
\label{sec:bubbles}
To convey surrounding avatars' location information (C2) to PVI without overwhelming them (C3), our design followed Hall's proxemic theory \cite{hall} to divide the virtual environment into different social spaces. Hall's proxemic theory correlated physical distances with social interactions that typically happen within that distance, such as the distance for intimate interactions versus the distance for friendly conversation. Three distances were defined: \textit{intimate distance (1 foot)} where people usually feel distress when this space is encroached upon unwillingly, \textit{personal distance (4 feet)} where people interact with familiar people, and \textit{social distance (12 feet)} where conversation with less familiar acquaintances or group happens. The space outside of the social distance is considered to be public space. 

\begin{figure}[!h]
    \centering
    \includegraphics[width=7cm]{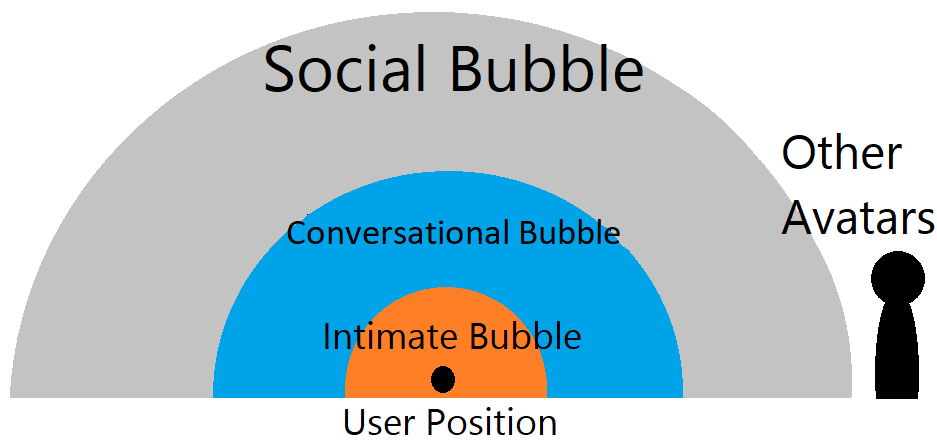}
    \caption{Conceptual diagram of bubbles.}
    \label{fig:bubbles1}
    \Description{This diagram is providing a visual example of how the bubbles can be imagined to look. The innermost is orange(Intimate bubble), with a blue middle shell(Conversation bubble), and finally a grey outer shell(Social bubble). The center represents the user's position.}
\end{figure}

We defined three virtual bubbles centered on a user based on the distance thresholds defined by Hall; the \emph{Intimate Bubble}, \emph{Conversation Bubble}, and \emph{Social Bubbles} (Figure \ref{fig:bubbles1}). We describe the social indication of each bubble:

\textit{\textbf{Intimate Bubble}} defines the space within the intimate distance. Avatars in this bubble signify that they are about to collide with the user's avatar. Since PVI cannot visually perceive whether they are too close to others or whether other people are invading their intimate space, it is important to alert them when avatars enter this bubble (C2).   

\textit{\textbf{Conversation Bubble}} represents the space between the intimate and personal distance. Avatars in this space are close enough to start a conversation with. Moreover, the user may need to pay attention to these avatars since they are close to overhear one's ongoing conversation. We thus generate audio feedback to notify the user if any avatar enters or exits this space (C2). 


\textit{\textbf{Social Bubble}} defines the space outside of the personal distance, but still within social distance. Avatars in this space are potential conversational partners, but would not be in the immediate distance to start the conversation. The user could then decide whether she is interested in approaching this person for a conversation. Compared to the Intimate and Conversation Bubble, we expect that avatar information in the Social Bubble would be less important. We thus design more subtle audio feedback to inform the user of the avatars in this bubble (C3). 


We do not consider avatars outside of the the social distance (i.e., public space in Hall's theory) since they are much less relevant to the users' current social context (C3).

\subsection{Audio Alternative Design via Iterations}
Based on the three bubbles, we designed spatial audio feedback to convey the surrounding avatars information, including the avatar identity and motion dynamics between these bubbles (C2), thus enabling PVI to build suitable social interactions upon sufficient avatar awareness. We also sought to make our audio feedback as least distracting and invasive as possible (C3). To achieve these goals, we designed and iterated on different audio feedback alternatives via a formative study. 

\subsubsection{\textbf{Formative Study}}
Following the method of user-centered design \cite{userdesign}, we conducted a formative study \cite{ji2022vrbubble} with six PVI (three male, two female, and one who preferred not to say) whose ages ranged from 22 to 58 ($mean=44.167$, $SD=13.348$). All participants were legally blind.

\textit{\textbf{Initial Design.}} In the formative study, we presented our initial design of VRBubble with the three bubbles described before. When an avatar entered or exited a bubble, the user heard an earcon with a verbal description of the avatar's information. Earcons were utilized as they are brief, abstract, and distinctive (C3) sounds that encodes particular information \cite{brewster1993evaluation}. We used different earcons to represents an avatar's moving dynamics between different bubbles. A two-beat earcon with increasing tone \upnotes{} (or decreasing tone \downnotes{}) was used to represent an avatar entering (or leaving) the Social Bubble; similar two-beat earcons with a different timbre were used for the Conversation Bubble; and an ``bumping'' sound earcon was used to indicate an avatar in the Intimate Bubble. All earcons were accompanied with a more informative verbal description (C2), reporting the avatar's name, relationship with the user (friend or stranger), and the relative position (``nearby'' for Conversation Bubble, ``In vicinity'' for Social Bubble). For example, the user heard ``Friend Alice nearby (or no longer nearby)'' if Alice's avatar entered (or left) the Conversation Bubble. All audio feedback were spatial audio rendered from the avatar's position. We prototyped VRBubble using web-based VR (C1, details in Section \ref{implemenation}). 

\textit{\textbf{Method and Findings.}} The formative study was hosted virtually through zoom and took roughly 2 hours for each participant to complete. Participants were given access to our VR environment through a url. Participants were introduced to the initial VRBubble design through a short tutorial. We then asked them to experience VRBubble in two tasks: a navigation task, and a conversation task. Both tasks were used to simulate common social contexts they could encounter in VR. During the prototype experiencing process, participants thought aloud, describing whether they like this feature or not and why. By the end, participants discussed how they wanted to improve VRBubble in different social context.

We summarize our major findings from the formative study. (1) While most participants found the earcon design useful, understanding the abstract earcons could create a steep learning curve for PVI. Some participants suggested more intuitive sound effects to present avatar information. (2) Verbal description is clear and easy to understand, but it could be distracting especially in a conversation context. Participants suggested shortening the verbal description to reduce distraction. (3) Participants had different preferences for audio feedback and valued various avatar information differently. Our design should provide the flexibility for users to customize their audio experience for different avatars and social contexts. 


\subsubsection{\textbf{Three Audio Alternatives}}
Based on the findings in the formative study, we designed three spatial audio feedback alternatives: earcons, verbal notifications, and real-world sound effects. Each alternative presented similar avatar information, including avatar identity (name and/or relationship with the PVI) and motion dynamics between bubbles (C2). 
We describe the design of three audio alternatives for each bubble:

\begin{figure}[htp]
    \centering
    \includegraphics[width=\linewidth]{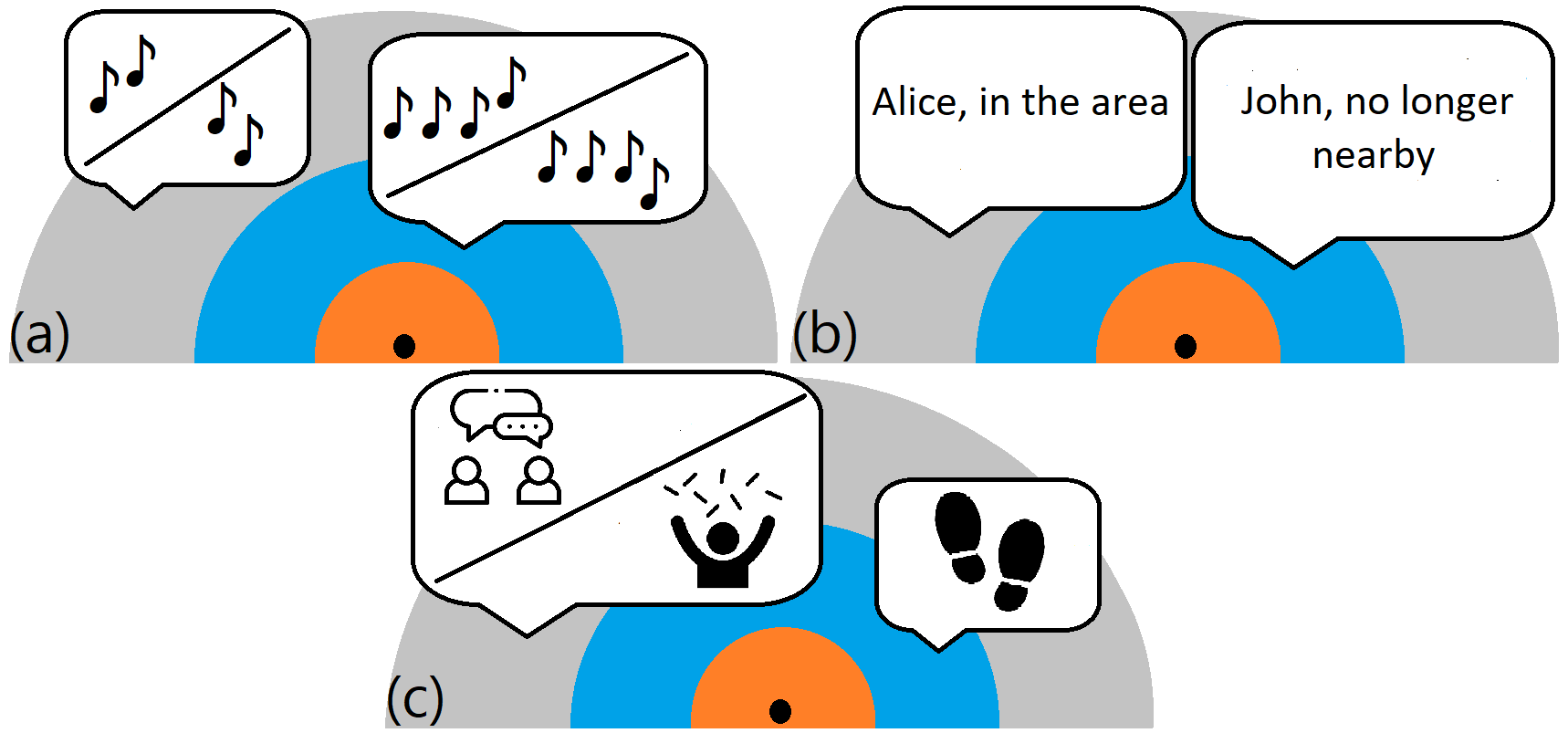}
    \caption{Three audio alternatives for different bubbles: (a) earcons, (b) verbal notifications, (c) real-world sound effects.}
    \label{fig:bubbles5}
    \Description{An illustration of each audio design. The leftmost provides a visual for how the earcon design is setup, with a two note earcon for the social bubble and a 4 note earcon for the conversation bubble. The center provides a visual for the verbal notification design for "Alice" entering the social bubble and "John" leaving the conversation bubble. The last visual shows the crowd and cheering sound in the social bubble and the footsteps sound in the conversation bubble.}
\end{figure}


\textit{\textbf{Earcon.}}
Given that earcon is abstract and has a high learning curve, we used earcons to present simple information, such as distinguishing different bubbles and different avatars (friend avatar vs. stranger avatar). Instead of completely different earcons, we designed associated earcons with distinctions to minimize the learning curve (C3). We used a two-beat earcon with the tone of the last beat increased \upnotes{} \footnote{Friend entering Social Bubble earcon:~\url{https://cdn.glitch.global/11db17e0-58dc-4f24-8390-bb2e597d1475/f_soc_in.mp3?v=1639640993569}} (or decreased \downnotes{} \footnote{Friend leaving Social Bubble earcon:~\url{https://cdn.glitch.global/11db17e0-58dc-4f24-8390-bb2e597d1475/f_soc_out.mp3?v=1639640998113}}) to indicate an avatar entering (or leaving) the Social Bubble. For the Conversation Bubble, since avatars in this bubble probably required more immediate attention, we used four-beat earcons with the tone of last beat increased \upsocial{} \footnote{Friend entering Conversation Bubble earcon:~\url{https://cdn.glitch.global/11db17e0-58dc-4f24-8390-bb2e597d1475/f_convo_in.mp3?v=1643983723059}}  or decreased \downsocial{} \footnote{Friend leaving Conversation Bubble earcon:~\url{https://cdn.glitch.global/11db17e0-58dc-4f24-8390-bb2e597d1475/f_convo_out.mp3?v=1643983757550}} to represent an avatar entering or leaving this bubble. To distinguish friend and stranger avatars, we adjusted the pitch and speed of the earcons, so that higher pitched, faster earcons indicated friends, while normal pitch and speed indicated strangers \footnote{Stranger entering Social Bubble earcon:~\url{https://cdn.glitch.global/11db17e0-58dc-4f24-8390-bb2e597d1475/s_soc_in.mp3?v=1649899925021}} \footnote{Stranger leaving Social Bubble earcon:~\url{https://cdn.glitch.global/11db17e0-58dc-4f24-8390-bb2e597d1475/s_soc_out.mp3?v=1639641013488}} \footnote{Stranger entering Conversation Bubble earcon:~\url{https://cdn.glitch.global/11db17e0-58dc-4f24-8390-bb2e597d1475/s_convo_in.mp3?v=1649900021917}} \footnote{Stranger leaving Conversation Bubble earcon:~\url{https://cdn.glitch.global/11db17e0-58dc-4f24-8390-bb2e597d1475/s_convo_out.mp3?v=1643983842245}}.
We used a game-like bump earcon \footnote{Intimate Bubble earcon:~\url{https://cdn.glitch.me/11db17e0-58dc-4f24-8390-bb2e597d1475\%2F483602__raclure__game-bump.mp3?v=1636388046675}} to signify when an avatar was in the Intimate Bubble.



\textit{\textbf{Verbal Notifications.}}
We provided clear and short (C3) verbal notifications to present avatar information. 
To reduce distraction, we informed avatar identity by only reporting the name. We no longer provided information on if the avatar was a friend or a stranger since in the real world a person would be able to know whether someone was a friend by name. We also used verbal notification to convey the avatar's general position based on their interaction with the bubbles. Specifically, we used ``in the area'' \footnote{Entering Social Bubble verbal:~\url{https://cdn.glitch.global/11db17e0-58dc-4f24-8390-bb2e597d1475/area.mp3?v=1643807674537}} and ``left the area'' \footnote{Leaving Social Bubble verbal:~\url{https://cdn.glitch.global/11db17e0-58dc-4f24-8390-bb2e597d1475/noarea.mp3?v=1643807963569}} to indicate when an avatar entered or left the Social Bubble, and ``nearby'' \footnote{Entering Conversation Bubble verbal:~\url{https://cdn.glitch.global/11db17e0-58dc-4f24-8390-bb2e597d1475/nearby.mp3?v=1643807938153}} and ``no longer nearby'' \footnote{Leaving Conversation Bubble verbal:~\url{https://cdn.glitch.global/11db17e0-58dc-4f24-8390-bb2e597d1475/nonearby.mp3?v=1643807989565}} for entering or leaving the Conversation Bubble. If an avatar, Alice, entered the Conversation Bubble, a user would hear ``Alice nearby.'' To signify that an avatar was in the Intimate Bubble, we used ``collided with'' \footnote{Intimate Bubble verbal:~\url{https://cdn.glitch.global/11db17e0-58dc-4f24-8390-bb2e597d1475/collidedwith.mp3?v=1643983652403}} followed by the avatar name.  

\textit{\textbf{Real-world Sound Effect.}}
This design alternative focused on providing realistic real-world sounds to intuitively support peripheral awareness and immersion, thus reducing the amount of effort needed to comprehend the audio notification (C3).  
We used crowd sound effects as the background sound, with different densities to represent the total number of avatars in all three bubbles within the social distance (12 feet). Two levels of crowd sound effects with increasing volume and densities indicate two levels of stranger avatar amount: 1-5 \footnote{Low ambient stranger crowd:~\url{https://cdn.glitch.me/11db17e0-58dc-4f24-8390-bb2e597d1475/stranger_crowd_small.wav?v=1639761138064}} and more than 5 \footnote{High ambient stranger crowd:~\url{https://cdn.glitch.me/11db17e0-58dc-4f24-8390-bb2e597d1475/stranger_crowd_medium.wav?v=1639761063563}} strangers.
Two levels of cheering crowd sounds were also used to highlight friend avatars due to their higher importance, indicating two levels of friend avatar amount: 1-5 \footnote{Low ambient friend crowd:~\url{https://cdn.glitch.me/11db17e0-58dc-4f24-8390-bb2e597d1475/friend_crowd_small.wav?v=1639761145228}} and more than 5 \footnote{High ambient friend crowd:~\url{https://cdn.glitch.me/11db17e0-58dc-4f24-8390-bb2e597d1475/friend_crowd_medium.mp3?v=1639761151401}} friends.
Moreover, we put more emphasis on the avatars in the Conversation Bubble since the user was more likely to converse with them. 
A spatial footstep sound effect was regularly played for every avatar within the Conversation Bubble. We generated different footsteps sounds to distinguish friend avatars from strangers, with the friend avatars presenting a higher pitch heel sound \footnote{Friend footstep:~\url{https://cdn.glitch.global/11db17e0-58dc-4f24-8390-bb2e597d1475/fstep.mp3?v=1643754264523}} and the stranger avatars presenting a boot stomp \footnote{Stranger footstep:~\url{https://cdn.glitch.global/11db17e0-58dc-4f24-8390-bb2e597d1475/sstep.mp3?v=1643754271035}}. We also generated a realistic bump sound \footnote{Collision:~\url{https://cdn.glitch.global/11db17e0-58dc-4f24-8390-bb2e597d1475/mixkit-quest-game-heavy-stomp-v-3049.wav?v=1644177861935}} to indicate an avatar entering the Intimate Bubble. 
\subsection{\textbf{Customization}}
Customization was key to our design since different PVI have different audio preferences and prioritize different information and goals. Instead of generating all audio feedback together (as we did in the initial design), we allowed users to select and combine different audio alternatives for different bubbles, avatars (e.g., friend vs. stranger), and social context to customize their experiences. For example, a user could choose verbal notifications for friends but earcons for strangers in the Social Bubble, and combine both audio alternatives for both friends and strangers in the Conversation Bubble. The user was also allowed to select none audio feedback for a specific type of avatar or bubble (e.g., stranger avatars in the Social Bubble).


\section{Evaluation}
We evaluated VRBubble with 12 PVI in different social contexts. We aimed to understand: (1) How effective is VRBubble in enhancing PVI's peripheral awareness of avatars in different social contexts? (2) How distracting is VRBubble? (3) How do PVI customize their audio experience for different avatars, bubbles, and social contexts? 
\subsection{Method}
\subsubsection{\textbf{Participants}}
We recruited 12 participants with visual impairments (7 male, 5 female) whose ages ranged from 20 to 68 ($mean=38.92$, $SD=13.79$, Table \ref{tab:demographic}). No participants were in the formative study. The participants were recruited through the National Federation of the Blind. We used a survey to check the participants' eligibility. Participants were eligible for our study if they were at least 18 years old, legally blind, and capable of independent consent. We also asked about participants' VR experience in the survey. We prioritized participants who had experience with VR or virtual environments. Among the 12 participants, nine had VR experience, such as audio-based VR games, VR for vision therapy, and VR environment exploration. However, no participants had experience with social VR. Participants were compensated at the rate of \$20 per hour.

\begin{table}[ht]
  \caption{Demographic information of 12 participants.}
  \centering
  \label{tab:demographic}
  \begin{tabular}{p{0.5cm} p{1cm} p{5.7cm}}
    \toprule
    ID & Age/Sex & Prior VR Experience\\
    \midrule
    P1 & 31/M & Vision therapy\\
    P2 & 57/M & 3D virtual home design game\\
    P3 & 37/M & Audio-based VR games; VR studies\\
    P4 & 68/M & VR environments; VR studies\\
    P5 & 44/F & Simulated street navigation VR app\\
    P6 & 34/F & VR headset games\\
    P7 & 49/F & Oculus Quest accessibility testing; VR games\\
    P8 & 25/F & None\\ 
    P9 & 33/F & None\\
    P10 & 46/M & PVI accessible audio games\\
    P11 & 23/M & Navigational VR study\\
    P12 & 20/M & None\\
  \bottomrule
  \Description{Participant information listed in order of : Participant, Age/Sex, Prior VR Experience.
  P1, 31/M, Vision therapy
  P2, 57/M, 3D virtual home design game
  P3, 37/M, Audio-based VR games; VR studies
  P4, 68/M, VR environments; VR studies
  P5, 44/F, Simulated street navigation VR app
  P6, 34/F, VR headset games
  P7, 49/F, Oculus Quest accessibility testing; VR games
  P8, 25/F, None
  P9, 33/F, None
  P10, 46/M, PVI accessible audio games
  P11, 23/M, Navigational VR study
  P12, 20/M, None
  }
\end{tabular}
\vspace{-2ex}
\end{table}

\subsubsection{\textbf{Apparatus.}}
\label{implemenation}
We built a custom VR environment with avatars and implemented both VRBubble and a baseline feature in the VR environment. To evaluate VRBubble in different contexts, we implemented two VR scenarios, a navigation scenario and a conversation scenario. We describe the design and implementation of our study environment.

\textit{\textbf{Baseline.}} We compared VRBubble's performance to a baseline feature. Our baseline was an audio beacon that consisted of a spatial beeping sound, in intervals, attached to each avatar (except for the participant avatar) in the environment. We chose this baseline since audio beacon was an effective and common way of conveying the presence of people and objects for PVI in the real world \cite{beacon}. Some current assistive technology for PVI (e.g., Microsoft Soundscape \cite{soundscape}) also uses audio beacon to label target destinations for PVI.  

\begin{figure}[t]
    \centering
    \includegraphics[width=\linewidth]{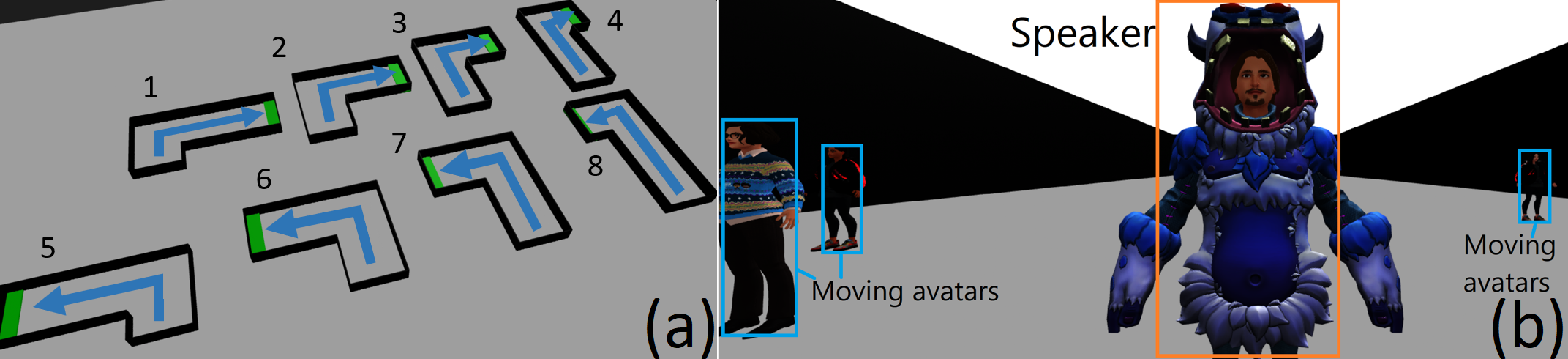}
    \caption{Two tasks: (a) navigation, (b) conversation.}
    \label{fig:clients}
    \Description{Visualizations of formative study tasks are shown here. On the left, (a), is the navigation task, with an overhead view of each of the 8 possible paths a participant would need to complete. Each path is the same overall length, with variations in which direction it turns and in what direction. On the right is the conversation context, with the speaker in the center and other avatars in the periphery, and each avatar is highlighted and labeled as the speaker or passing-by avatars.}
\end{figure}

\textit{\textbf{VR environment and Two Scenarios.}} We built a custom VR environment with two key VR scenarios: a navigation scenario and a conversation scenario. In the navigation scenario, we generated a VR space plan with eight similar navigation routes (Figure \ref{fig:clients}(a)). All routes were approximately 60 ft with one turn (either turn left or turn right). All routes were considered to be the same according to Ishikawa et al.'s study design \cite{ishikawa2008wayfinding} since they had same length and same number of turns. Avatars were rendered at along the routes at various distances (Figure \ref{fig:clients}(a)). 

In the conversation scenario, we rendered one avatar in front of the user, conversing with the user. Other avatars were rendered and moving around in the virtual space during the conversation (Figure \ref{fig:clients}(b)). 

\textit{\textbf{Features for Basic Movements}}\label{BasicMovement}
To enable participants to navigate in the virtual environment and experience the avatar awareness features, we added two fundamental features to allow PVI to move and turn in a virtual environment: (1) \textit{Movement:} PVI can control the movement of their own avatar via arrow keys, for example, up arrow to move forward and left arrow to move left. Each key press would move the PVI a foot in the virtual space and play a footstep sound to notify the PVI. This footstep sound \footnote{User footstep:~\url{https://cdn.glitch.me/11db17e0-58dc-4f24-8390-bb2e597d1475\%2F571698__rainial-co__cotton-thud-7.wav?v=1636411987805}} is centered on the user's avatar and differs from the footsteps used for VRBubble by pitch. (2) \textit{Turning.} A user can turn her avatar to the left (or right) by pressing the left (or right) arrow keys while holding the shift key. A ticking sound \footnote{Turn audio:~\url{https://cdn.glitch.me/11db17e0-58dc-4f24-8390-bb2e597d1475\%2F268108__nenadsimic__button-tick.wav?v=1636008594574}} would play for each 90 degree turn, providing a sense of how much the PVI had turned.

\textit{\textbf{VR Environment Implementation and Setup.}} We implemented the VR environment and all the features with A-Frame \cite{aframe}, a web framework that generates 3D spaces via HTML and Javascript. We hosted our prototype on Glitch \cite{glitch}, a browser-based platform that allows users to host and share applications as well as share the source code, to enable a remote setup for the study. Participants and researchers can go to the same web address to join the same VR environment. Once connected, researchers can input unique command sequences via a keyboard to adjust the participant's client, such as adjusting the audio alternatives in VRBubble or moving the participant's avatar to a different VR scenario. Participants' behavioral data (e.g., their positions at different times) was then logged and output into the researchers' browser console. 

We asked participants to share their screen with audio via Zoom to ensure the commands from the researchers were properly reflected on the participants' end and to confirm that the audio feedback was properly generated at the participants' end.

\subsubsection{\textbf{Procedure.}} Our study consisted of a single session that lasted 2 hours for each participant. We conducted the study remotely via a Zoom video call. An initial interview was first conducted to ask about participants' prior experience with VR, especially social VR. Then we asked about their social experiences in real life. 

We then asked participants to experience our VR environment with the awareness features. Participants were requested to use their earphones for the study, since we generate spatial audio. We sent a url to our VR environment through email before the study. If any participant had difficulty accessing their email, we sent the link again through the Zoom chat or read the link out-loud to them. 

After the participant successfully joined the VR environment, we evaluated participants' experience and performance in two context: (1) a navigation context and (2) a conversation context. In each context, participants completed navigation or conversation tasks with two awareness features---VRBubble and the baseline (i.e., audio beacon). We counterbalanced the awareness feature and the context across all participants. We elaborate the details of the remaining study in three phases: tutorial, navigation, an conversation.

\textbf{\textit{Tutorial: Introducing the Features.}}  
We conducted a tutorial session to teach participants how to use our VR prototypes. We first introduced the interactions for basic movements (arrows keys to move and shift + left/right arrow key to turn, as explained in Section \ref{BasicMovement}). We then rendered an empty VR space and encouraged participants to move around until they were familiar with the controls and the audio feedback produced by the avatar movement. We then introduced VRBubble and the baseline. 

In the tutorial for VRBubble, we introduced each audio alternative one by one. For each alternative, we demonstrated the audio feedback of both friend avatar and stranger avatar for each Bubble. We then rendered avatars in the VR space and asked participants to freely explore the environment with this audio alternative until they felt fully familiar with it. During the exploration, participants were asked to think aloud, talking about their experiences with this audio design, including whether they like this design or not and the reason. We counterbalanced the order of the audio alternatives using Latin Square \cite{Nahler2009latin}. We used the similar method to introduce the baseline.

\textbf{\textit{Navigation.}}
We evaluated the effectiveness of VRBubble in the navigation context. Each participant conducted 15 trials of navigation tasks along the routes in Figure \ref{fig:clients}a in different conditions. We randomly selected a route for each trial. To enable participants to follow the route and arrive at the destination, we generated audio-based verbal instructions. Participants heard ``turn left (or right)'' when they needed to turn, and heard ``arrived destination'' when they reached the end of the route. For each trial, the participant was first moved to the start point of a route. When the research announced ``Start'', they would navigate by following the verbal instruction until arriving at the destination. We asked participants to complete each navigation task as quickly as possible.

Participants first conducted three trials of basic navigation tasks without any avatars around. These trials helped participants get familiar with the navigation tasks. We then asked participants to customize their audio feedback in VRBubble based on the navigation task. To guide participants through the customization, we asked them to choose and combine audio feedback for the five following situations: (1) when a friend enters or leaves the Social Bubble; (2) when a stranger enters or leaves the Social Bubble; (3) when a stranger enters or leaves the Conversation Bubble; (4) when a friend enters or leaves the Conversation Bubble; (5) when an avatar collides with the user in the Intimate Bubble. For each situation, participants could select any combination of audio feedback ranging from multiple to none at all (Figure \ref{fig:customizetable}). During the customization process, the researcher adjusted the audio alternatives in real time for the participant to experience the selected feedback and make immediate corrections. After customizing VRBubble, we asked about the reason behind their choices and any suggestions they have for customizing. Once the customization phase was completed, the researcher adjusted the audio feedback in VRBubble to the participant's preferred combination. 

After the customization, participants continued to perform another 12 trials of navigation tasks with avatars around. Participants were asked to complete two tasks at the same time: a primary navigation task (i.e., navigating to the destination following the verbal instructions) and a secondary avatar awareness task (i.e., memorizing how many avatars they've passed by and who they are). We asked participants to reach the destination as quickly as possible, but also try their best to perceive and memorize the surrounding avatars. Before the navigation task, we informed the participants ahead that we would asked them to report the avatar number and names they remembered after each trial. However, if the task was too difficult, we asked participants to prioritize the primary navigation task. 

Participants completed above tasks in two awareness conditions: using VRBubble with the customized audio alternatives, and using the baseline. Each condition included six trials of navigation. To further understand the effectiveness of VRBubble in situations with different avatar complexity, we generated avatars in two amount levels: Low Amount (1-5 avatars) and High Amount (6-10 avatars). We decided whether a trial had low or high amount of avatars randomly but ensured that participants experienced each avatar amount level for three trials per awareness condition. For each trial, we randomly generated avatars within the corresponding avatar amount range, and the avatars were randomly distributed along the navigation route. We guaranteed that all avatars were within the social distance (12 feet) to the route, so that all avatars were detectable by VRBubble. After each navigation trial, we asked participants to report their perceived avatar information, including the total avatar number, the number of friend avatars, the number of stranger avatars, and the avatar names they heard. We timed participants' navigation time for all trials. 
 
By the end, we asked about participants' general experience with VRBubble and the baseline. Participants rated the effectiveness, distraction, and immersion of both features via 5-point likert scale scores.   
Participants also discussed their feedback and suggestions for improving VRBubble in the navigation context.

\textbf{\textit{Conversation.}} The procedure of the conversation task was similar to the navigation task. This phase of the study was conducted in the conversation scenario (Figure \ref{fig:clients}b). Participants conducted three trials of conversation task in different conditions. For each trial, the participant faced an avatar who asked 10 yes/no questions. Participants were asked to answer all questions as quickly and accurately as possible. A researcher on the team acted as the conversation avatar to ask the questions. The next question was asked immediately after the participant answered the prior question. 

Participants first conducted one trial of ``dry-run'' task, answering 10 questions without any distraction to get familiar with the task. We then asked them to customize the VRBubble feedback again for the conversation context. Participants then continued to perform two trials of conversation tasks with avatars around. One trial used the customized VRBubble and the other used the baseline. For each trial, participants conducted two tasks simultaneously: a primary conversation task (i.e., answering 10 yes/no questions) and a secondary awareness task (i.e., perceiving and memorizing surrounding avatars). The primary task was prioritized if the task was too difficult. After each trial, we asked participants to recall the avatar information, including total avatar number, the number of friends and strangers, and the avatar names. 

To balance the avatar complexity in each trial, we pre-defined two similar groups of avatar moving dynamics. Both groups had eight avatars spawn in and move around the user over the duration of 70 seconds during the conversation process. Each group had four friend avatars and four stranger avatars. However, the order of avatars and the timing they spawn in were different between the two groups. We counterbalanced the avatar groups with the awareness feature used across participants. 

We pre-planned 30 yes/no questions. All questions were easy-to-answer questions that asked about a specific fact about the participants to minimize the variances caused by participants' literacy and cognitive ability. Some example questions included ``Have you ever been to Italy?'', ``Are you a coffee drinker?'' We randomly pooled 10 questions for each trial and also made sure not to repeat any questions across the study for each participant. We measured participants' response time for each question. 

After all trials, we asked all 30 yes/no questions again without any distractions and collected participants' answers as the correct answers. Participants then assessed the effectiveness, distraction, and immersion of VRBubble and the baseline for the conversation context via 5-point likert scale scores. Finally, participants talked about their qualitative feedback and suggestions for VRBubble in the conversation context.

We ended the study by discussing with the participants about their general experiences with VRBubble compared to the baseline, as well as their desired awareness technology for different contexts in social VR in the future. 

\subsubsection{\textbf{Analysis}}
We detail our analysis methods for both the quantitative and qualitative data in our study.

\textbf{\textit{Navigation Performance}}
We first analyzed the impact of VRBubble on participants' performance in the navigation task, including navigation time and the error rate when recalling the number of avatars they passed by. We had three within-subject factors: \textit{Feature} (two levels: \textit{VRBubble}, \textit{Baseline}) that defined the avatar awareness feature participants used, \textit{AvatarAmount} (\textit{Low: 1-5}, \textit{High: 6-10}) that specified the amount of avatars in the space, and \textit{Trial} (1-3) that defined each navigation task in a specific condition. 
We had two measures: \textit{NavigationTime}---the time taken in seconds by the participant to reach the end of the navigation path, and \textit{AvatarErrorRate}---the ratio of the difference between the reported avatar number and the actual number to the actual number of avatars.

We also added a between-subject factor, \textit{Order} (\textit{VRBubble-Baseline}, \textit{Baseline-VRBubble}) to our model for counterbalance validation. The Shapiro-Wilk test showed that both NavigationTime ($W=.867, p<.001$) and AvatarErrorRate ($W=.898, p<.001$) deviated significantly from a normal distribution, so we used Aligned Rank Transform for nonparametric factorial ANOVAs (ART) \cite{kay2021artool} \footnote{ART is a non-parametric approach to factorial ANOVA or linear mixed effect model} to model our data. We found no significant effect of Order ($F_{1,10}=2.296, p=.161$) on the navigation time and no significant effect of the interaction between Order and other factors on navigation time. Similarly, we also found no significant effects of Order ($F_{1,10}=.0005, p=.982$) on avatar error rate and no significant effect of the interaction between Order and other factors on error rate.


\textbf{\textit{Conversation Performance}}
We then analyzed the impact of VRBubble on participants' performance in the conversation task, including the response time per question, accuracy of answering questions, and the error rate when recalling the number of passed-by avatars during the conversation. We had two within-subject factors: \textit{Feature} (\textit{VRBubble}, \textit{Baseline}), and \textit{AvatarGroup} (\textit{Group1}, \textit{Group2}) that indicated which pre-defined set of avatar dynamics was used during each conversation trial.
We had three measures: \textit{AvatarErrorRate}, \textit{ResponseTime}---the mean response time for the participant to answer a question, and \textit{IncorrectAnswers}---the number of questions that participants answered inconsistently compared to the standard answers or questions that needed to be repeated.

A between-subject factor, \textit{Order} (\textit{VRBubble-Baseline}, \textit{Baseline-VRBubble}), was added to valid our counterbalancing. The Shapiro-Wilk test showed that ResponseTime ($W=.886, p=.011$) and IncorrectAnswers ($W=.654, p<.001$) were non-normally distributed , while AvatarErrorRate ($W=.972, p=.708$) followed a normal distribution. We thus used ART to model ResponseTime and IncorrectAnswers, and used ANOVA \cite{girden1992anova} for AvatarErrorRate. We found no significant effect of Order on the response time ($F_{1,8}=.293, p=.603$) or questions answered incorrectly ($F_{1,8}=.802, p=.397$) and no significant effect of the interactions between Order and other factors on response time or questions answered incorrectly. Similarly, we found no significant effects of Order ($F_{1,8}=.702, p=.426$) or its interactions on the error rate of recalling avatars.

Moreover, we found no significant effect of AvatarGroup or its interactions on ResponseTime ($F_{1,8}=.026 , p=.876$), IncorrectAnswers ($F_{1,8}=.82 , p=.392$), and AvatarErrorRate ($F_{1,8}=.157 , p=.097$). These results confirmed that the two pre-defined sets of avatars had similar complexity. 

\textbf{\textit{Post hoc Analysis.}} Based on the results of the ART and ANOVA models, we conducted post hoc tests to further investigate the relationship between the different levels of significant factors. Specifically, we used paired t-test for normally distributed data, and used the pairwise Wilcoxon signed-rank test for non-normally distributed data. We corrected the p-value threshold with Bonferroni Correction.

\textit{\textbf{Qualitative Analysis.}}
We video recorded the whole study and transcribed participants' verbal feedback and responses to the interview questions with an automatic transcription service, Otter.ai. Researchers on the team also read through the transcripts and corrected the auto transcription errors in the transcripts. We analyzed the transcripts using the the qualitative analysis method described by Salda{\~n}a \cite{saldana2021coding}. Two researchers first independently coded the first four participants' transcripts. They then discussed and generated a codebook upon agreement. One researcher then followed the codebook and coded the rest of the data. When a new code emerged, the two researchers discussed again and updated the codebook. The codes were then organized and categorized into themes using affinity diagram. 

\subsection{Results}
We report the impact of VRBubble on participants' performance in different context, including both a navigation and a conversation task. We also report participants' experiences with different audio alternatives and their preferences when customizing VRBubble for different context.

\subsubsection{VRBubble in Navigation}
For the navigation task, we found that VRBubble provided the user with more detailed and accurate information about surrounding avatars, compared to the baseline. However, we also noticed that VRBubble could be more distracting than the baseline. We report the detailed results below. 

\textit{\textbf{Avatar Amount Estimation.}}
We analyzed the effectiveness of VRBubble on enhancing PVI's awareness of the amount of avatars nearby in a navigation context. An ART analysis showed a significant effect of the awareness feature on participants' error rate when estimating the amount of avatars they passed by ($F_{1,110}=33.54, p<.001$). We then conducted a \textit{post hoc} paired Wilcoxon signed-rank test and found that participants had a significantly lower error rate when using VRBubble than the baseline ($V=1566.5 , p<.001$, error rate with VRBubble: $mean=.239, SD=.26$, baseline: $mean=.465 , SD=.323$). This result demonstrated that VRBubble significantly improved PVI's awareness of the surrounding avatar amount. 


%
We examined the error rate more closely for environments with different avatar complexity (AvatarAmount): low amount of avatars vs high amount of avatars. A \textit{post hoc} Wilcoxon signed-rank test showed that VRBubble significantly decreased participants' error rate of avatar number estimation regardless the avatar amount was high ($V=585 , p<.001$) or low ($V=259, p=.01<.025$ with Bonferroni Correction). 
This indicated that the effectiveness of VRBubble was consistent across environments with different avatar complexity.

Participants' qualitative feedback also explained the effectiveness of VRBubble. Ten participants expressed positive sentiment towards VRBubble, emphasizing its ability to provide distinguishable audio feedback for each avatar, with seven noting VRBubble created a better sense of distance for avatars via the different bubbles. In contrast, participants found the audio beacon baseline hard to use since all avatars shared the same beacon sound and it was difficult to distinguish different avatars, especially when avatars were close together. P6 \textit{``had a hard time differentiating if it was the beacon for the additional avatar or the same avatar. I had a hard time distinguishing how many people or avatars there were.''}

\textit{\textbf{Avatar Identification.}}
Compared to the baseline, VRBubble enabled participants to collect more detailed information about surrounding avatars, including avatar names and the relationship to the user.


When customizing VRBubble, nine participants selected verbal notification that reported avatar names. With this feature, participants in general recalled the names of passed-by avatars accurately with a mean accuracy of 0.715 ($SD=.305$). Six participants expressly liked VRBubble's capacity to provide information about each individual avatar. For example, P3 emphasized the importance of name identification in engaging in social interactions: \textit{``[The name] would potentially determine whether or not I would want to go and speak to them ... or even somebody I want to avoid ... I can see that having been named there is a useful feature.''} However, we noticed that the avatar identification accuracy of some participants dropped to below 0.5 in more crowded avatar environments. 
This was because the verbal notifications from different avatars would overlap and become confusing when the avatar reached high amounts. 
As P5 noted, \textit{``If it's not too many people happening at once then I can feel it out, but then when so many people are happening, then I can't really figure it out.''}

Ten participants' VRBubble customization supported friend identification (i.e., earcon or real-world sound effect for Conversation Bubble). Most participants distinguished friends and strangers accurately with a mean accuracy of 0.658 for friends ($SD=.367$) and 0.656 for strangers ($SD=.376$). 
Ten participants agreed that discerning between friends and strangers was important for them to decide what social actions to take. As P2 said, \textit{``I might want to ask a stranger what time it is. But if I hear friendly footsteps coming towards me, I want to know who's approaching and I can strike up conversation.''} However, P4 was an outlier who consistently reported many more avatars than the actual friend and stranger numbers in the space. This was because P4 used the abstract earcons for the Conversation Bubble and found it difficult to discern the pitch changes between friends and strangers. 


The two participants (P5, P10) chose verbal notification for the Conversation Bubble and thus could not distinguish friends and strangers. They customized VRBubble based on their real-life experiences, as in real-life they would recognize if someone was a friend or stranger by name. We envisioned that users would be able to distinguish friends and strangers via verbally reported names in a real social VR setting after long-term use. 



\textit{\textbf{Impact on Navigation Time.}}
We compared the effects of awareness feature on participants' navigation time. With an ART model, we found a significant effect of awareness feature on navigation time ($F_1=36.708 , p<.001$). A post hoc paired Wilcoxon signed-rank test with correction showed that participants navigated significantly slower when using VRBubble than using the beacon baseline ($V=482 , p<.001$, navigation time with VRBubble: $mean=36.507 , SD=14.897$, baseline: $mean=31.293 , SD=11.033$). 
The result indicated that VRBubble distracted participants more than the baseline in the navigation task. 



We further investigated the effect of VRBubble under different avatar complexity (low amount of avatars vs. high amount). We found a significant effect of the interaction between Feature and AvatarAmount on participants' navigation time ($F_1=13.676 , p<.001$). Using a \textit{post hoc} Wilcoxon signed-rank test with Bonferroni Correction, we found that, there's no significant difference in time between VRBubble and baseline when the avatar number was low ($V=209 , p=.052$), but participants walked significantly slower with VRBubble than the baseline when the avatar number was high ($V=66, p<.001$). This indicated that VRBubble was only more distracting when the environments became crowded. 

The distraction of VRBubble could be caused by the more diverse audio feedback provided by VRBubble, where participants spent longer time to associate the sounds with different events. Six participants mentioned being distracted and slowed down since they tried to remember the meaning of the different sounds.
As P5 explained, \textit{``If I go too fast, I might miss it ... I have to pay attention on top of all the other avatars leaving, coming in, and coming out.''}

Another reason that may cause a slower navigation was participants' interest. Since VRBubble enabled participants to receive more information about the surrounding avatars, it aroused participants' interest to explore the environment during the navigation. In our study, while we emphasized that navigation was the primary task, we still observed that some participants (e.g., P2, P5, P12) stopped, looked around, and even moved their avatar left and right to explore nearby avatars with VRBubble. In contrast, since the baseline did not provide much information, participants mostly focused on the navigation.



\subsubsection{VRBubble in Conversation}
Unlike in the navigation task, VRBubble generally did not perform significantly different from the baseline in the conversation context. We report the detailed measures below.

\textit{\textbf{Avatar Amount Estimation.}}
We analyzed the effectiveness of VRBubble on enhancing PVI's awareness of avatar amount nearby in a conversation context. An ANOVA analysis showed no significant effect of the awareness feature on participants' error rate when estimating the amount of avatar they passed by ($F_{1,8}=1.431, p=.266$). 
Compared to the navigation task, we found that participants made more errors when estimating the avatar number with VRBubble in the conversation task ($mean=.305 , SD=.191$). The error rates went especially high when there are more avatars around. Interestingly, we noticed that the reported avatar number by participants never exceeded five across all trials, even though there were usually more avatars present. Specifically, eight participants encountered more than five avatars in the conversation task, and the highest avatar number was eight. This might indicate a limit in how much audio feedback can be peripherally processed (or a upper bound of avatar number that can be perceived) while focusing on an audio-centered task such as conversation. As P4 noted, \textit{``Because you had to answer the question, so you couldn't put all your brainpower into [the surrounding avatars]. I just don't think the human brain is segmented enough to handle both those tasks at the same time.''}

\textit{\textbf{Avatar Identification.}}
Similar to the navigation task, participants appreciated VRBubble's ability to identify surrounding avatars in real-time since it enabled them to dynamically adapt to the social environments. Ten participants indicated that knowing who was in the immediate surroundings had an impact on their social behaviours and what they would say during a conversation.

Seven participants customized VRBubble to allow for name identification. However, participants' avatar recognition accuracy were lower ($mean=.655 , SD=.282$) compared to the navigation task ($mean=.715, SD=.305$). Moreover, ten participants customized VRBubble to distinguish friends from strangers. Similarly, they were less accurate compared to those in the navigation task, both when identifying friends ($mean=.567 , SD=.323$) and when identifying strangers ($mean=.467 , SD=.302$). The lower accuracy could be attributed to the more attention-demanding nature of the conversation task, and the direct conflict between verbal notifications and the conversation.

\textit{\textbf{Impact on Conversation.}}
We investigated the impact of VRBubble on participants' ability to answer questions in a conversation. With an ART model, we found no significant effect of the awareness feature on the number of questions answered incorrectly ($F_{1,8}=.647 , p=.444$). Most participants answered all questions correctly with both features. Four participants (P2, P4, P5, P12) presented incorrect answers while using VRBubble, and four participants (P4, P7, P10, P11) did so with the baseline. P2 was the only participant who answered two questions incorrectly, while everyone else answered one question incorrectly. 

Moreover, we compared the effects of VRBubble and the baseline on participants' response time to questions in the conversation. With an ART model, we found no significant effect of Feature on response time ($F_{1,8}=.206 , p=.662$).
These results showed that VRBubble and baseline were generally at the same distraction level in the conversation task.

\subsubsection{Perceived Experience with VRBubble.}
\begin{figure}[htp]
    \centering
    \includegraphics[width=8cm]{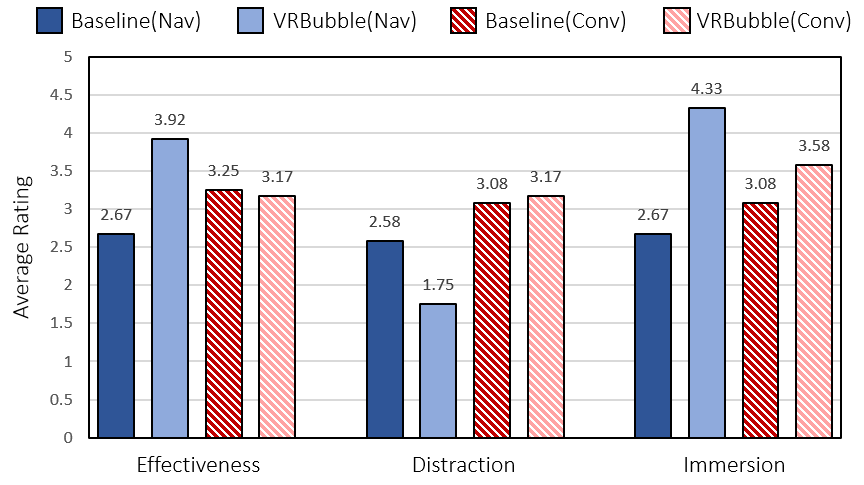}
    \caption{Likert scale score averages for each task: navigation Task (nav) and conversation task (conv).}
    \label{fig:bar}
    \Description{From left to right:
    For effectiveness; baseline navigation (2.67), VRBubble navigation (3.92), baseline conversation (3.25), VRBubble conversation (3.17).
    For distraction; baseline navigation (2.58), VRBubble navigation (1.75), baseline conversation (3.08), VRBubble conversation (3.17).
    For immersion; baseline navigation (2.67), VRBubble navigation (4.33), baseline conversation (3.08), VRBubble conversation (3.58).}
\end{figure}

We compared participants subjective experience with VRBubble and the baseline from three dimensions, including effectiveness, distraction, and immersion. Figure \ref{fig:bar} shows the mean scores given by participants for each dimension. Participants perceived VRBubble ($mean=3.917 , SD=.9$) to be more effective than the baseline ($mean=2.667 , SD=1.231$) at enhancing avatar awareness in the navigation task, while perceiving VRBubble ($mean=3.167 , SD=1.403$) to be slightly less effective than the baseline ($mean=3.25 , SD=1.485$) for the conversation task. These scores matched our expectations given that the conversation task was more attention-demanding and participants had less cognitive capacity to use VRBubble.  

Participants perceived more immersion in the virtual space while using VRBubble ($mean=4.333 , SD=.888$) than the baseline ($mean=2.667 , SD=1.557$) during the navigation task. They also felt more immersed during the conversation task when using VRBubble ($mean=3.583 ,  SD=1.443$) than when using the baseline ($mean=3.083 , SD=1.505$). Six participants found the audio beacon baseline to be annoying or distracting, and thus diminishing their immersive experience. In contrast, VRBubble provided a better sense of presence by dividing the social space and associating the avatar dynamics with social indications. As P3 mentioned, \textit{``The bubble system was much more immersive, just because it gives you a couple of clear lines between when you're entering and leaving someone's space. The beacon is just super impersonal.''}

Interestingly, participants perceived VRBubble ($mean=1.75 , SD=.754$) to be less distracting than the baseline ($mean=2.583 , SD= 1.379$) for the navigation task, despite taking significantly more time to complete the task with VRBubble. This further confirmed that the slower navigation could be caused by participant's increased interest in exploring the surrounding with VRBubble instead of distraction. 
For the conversation task, participants perceived VRBubble ($mean=3.167 , SD=1.467$) to be slightly more distracting than the baseline ($mean=3.083 , SD=1.443$). Five participants expressed that the amount of information from VRBubble caused distractions during conversation, suggesting simpler or reduced audio was preferred for audio-intensive tasks. 

\subsubsection{Experience with VRBubble and its Audio Alternatives}
All participants agreed that the Bubble concept based on social distances was novel and effective, and expressed favor that VRBubble provided them with more information than the baseline did. Seven participants liked that VRBubble separated the space and conveyed a sense of distance to the user. P3 noted, \textit{``I like the three different distances. If I walk into a social situation, I wish I had a bubble radar just to know when people are in my immediate vicinity versus people a little ways away. It is just nice to have a concept of relative [distance].''} Participants also liked the sense of avatar presence and movement that was conveyed by the feedback from the bubbles.
We then elaborate participants' experiences with different audio alternatives below. 

\textit{\textbf{Earcon.}}
Four participants liked the brevity and non-intrusiveness of the earcons. Unlike the continuous audio feedback in the baseline, participants liked that the earcons discretely notified them every time an avatar entered a bubble. For example, P8 had difficulty with discerning spatial audio. She reported, \textit{``I like how concrete [the earcon] is. I have a really hard time with spatial awareness. So, without necessarily having to keep track of the space because there are those different [earcons], it's very helpful.''} Three participants noted that earcons for each avatar event could be easily noticed and distinguished, which could be used to reduce audio overloading in crowded environments. As P4 indicated, \textit{``To signify the person is leaving, I'd want the [earcon] sound that shows the person's going out. Because then you don't have too many verbal cues, you know, to not listen too much.''}

The main deterrent of earcons was the deep learning curve. Six participants found it hard to recognize different earcon sounds. P4 pointed out that the meanings of earcons would have to become second nature in order to be effective in a dynamic social environment: \textit{``You don't want to have to think about it, you want to immediately react.''} Five participants had issues with discerning the pitch difference between friend and stranger earcons. As P1 said, \textit{``there were more and more unexpected [earcons] that threw me off and, being not as familiar with the tones, second guessing if [the earcon] was higher [pitched] or lower, entering or exiting.''} P4 suggested that instead of using different pitches, completely different earcons should be used for different types of avatars.

\textit{\textbf{Verbal Notification.}}
Nine participants liked that the verbal notifications were informative and easy to understand. They expressed a desire for sufficient information about their surroundings to adjust their behaviors, especially in the context of conversation. P3 emphasized that distinguishing between friends and strangers can help protect his privacy, \textit{``there may be something that you're willing to share with a friend, but not necessarily with a complete stranger.''}

The drawback of verbal notifications was its verbosity, which caused more audio overlapping between avatars. Nine participants mentioned this problem in the study. Four participants also noted that hearing a verbal notification immediately drew their attention, which could be distracting in certain scenarios, such as conversation. Due to these issues, three participants cared only about verbal notifications for important people, such as friends, rather than being verbally informed about every avatar.

\textit{\textbf{Real-world Sound Effect.}}
Six participants liked the footstep and crowd noises since they were ambient and non-distracting. The audio had no sudden alerts and was quiet enough to tune out when focusing on the primary task. The crowd sounds were also immersive and helped convey a general sense of the social environment, such as if the user was at an informal party or a business meeting. For example, P4 preferred using real-world sound effects in the conversation task for an immersive experience. As he expressed, \textit{``I was not gearing [my customization] toward the best [avatar recognition] results but [was] gearing it toward getting a real feel for it.''} P3 even suggested providing \textit{``different types of ambient background noises that meet the theme of whatever rooms.''} Two participants (P3, P6) also felt that the two levels of ambient crowd sound with different loudness was sufficient, and any further levels may be too loud or unable to contribute any additional useful information. Moreover, two participants (P1, P2) liked the spatial footstep audio in the Conversation Bubble, which helped them constantly track important avatars in the conversational distance.

However, six participants disliked the constant real-world sound effects since it was hard to for them to distinguish each avatar via the general crowd sounds. P6 thus suggested combining the crowd sound with an overall description of the avatar number. Two participants (P7, P10) had difficulty distinguishing friend and stranger avatars via the different footstep sounds, especially with other louder audio in the environment.

\subsubsection{Customization of Audio Alternatives}
All participants appreciated the flexibility of customizing the audio feedback in VRBubble. For example, P8 had an auditory processing disorder that caused large quantities of sound to be difficult to process. As she mentioned, \textit{``[I like] the idea that you can customize how you're going to handle an event or have part of an event made accessible to you via sound. It is not something most people think about and I really appreciate it.''} 

\begin{figure}[htp]
    \centering
    \includegraphics[width=8cm]{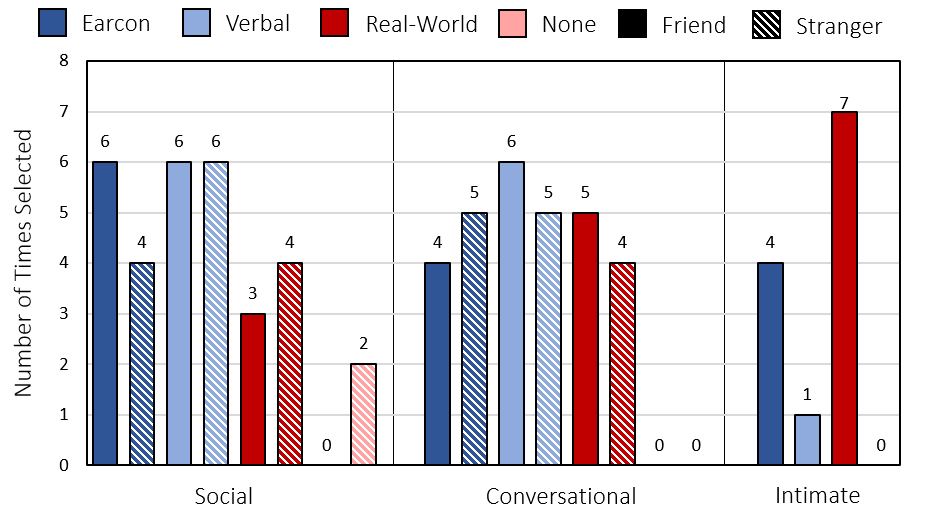}
    \caption{VRBubble customization for navigation task: Side-by-side comparison between amount of times that each audio alternative was selected for friends and strangers, at each bubble. Intimate bubble is shared between friends and strangers.}
    \label{fig:navcustomize}
    \Description{Bar chart of number of times an audio alternative was selected. From left to right: For the social bubble; friend earcon (6), stranger earcon (4), friend verbal (6), stranger verbal (6), friend real-world (3), stranger real-world (4), friend none (0), stranger none (2). For the conversational bubble; friend earcon (4), stranger earcon (5), friend verbal (6), stranger verbal (5), friend real-world (5), stranger real-world (4), friend none (0), stranger none (0). For the intimate bubble; earcon (4), verbal (1), real-world (7), none (0)}
\end{figure}

\begin{figure}[htp]
    \centering
    \includegraphics[width=8cm]{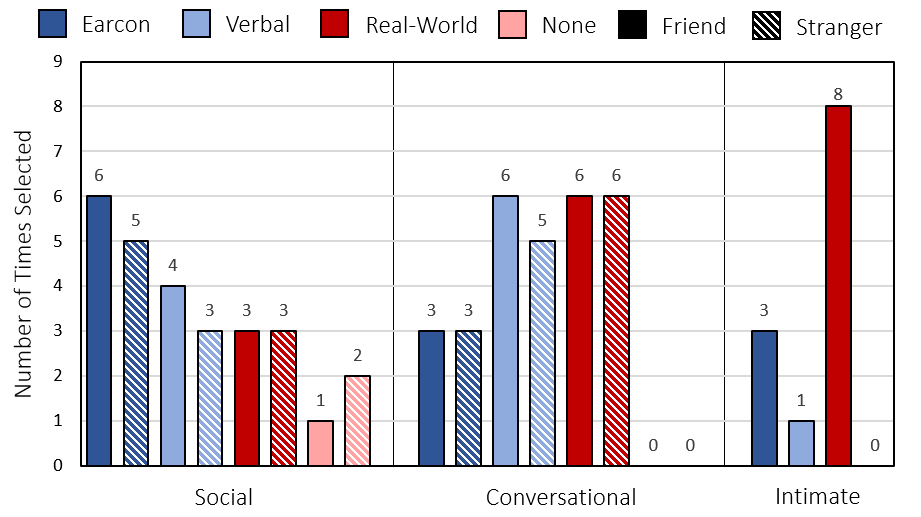}
    \caption{VRBubble customization for the conversation task: side-by-side comparison between amount of times audio alternative is selected for friends and strangers, at each bubble. Intimate bubble is shared between friends and strangers.}
    \label{fig:convocustomize}
    \Description{Bar chart of number of times an audio alternative was selected. From left to right: For the social bubble; friend earcon (6), stranger earcon (5), friend verbal (4), stranger verbal (3), friend real-world (3), stranger real-world (3), friend none (1), stranger none (2). For the conversational bubble; friend earcon (3), stranger earcon (3), friend verbal (6), stranger verbal (5), friend real-world (6), stranger real-world (6), friend none (0), stranger none (0). For the intimate bubble; earcon (3), verbal (1), real-world (8), none (0)}
\end{figure}

\begin{figure*}[!htp]
    \centering
    \includegraphics[width=15cm]{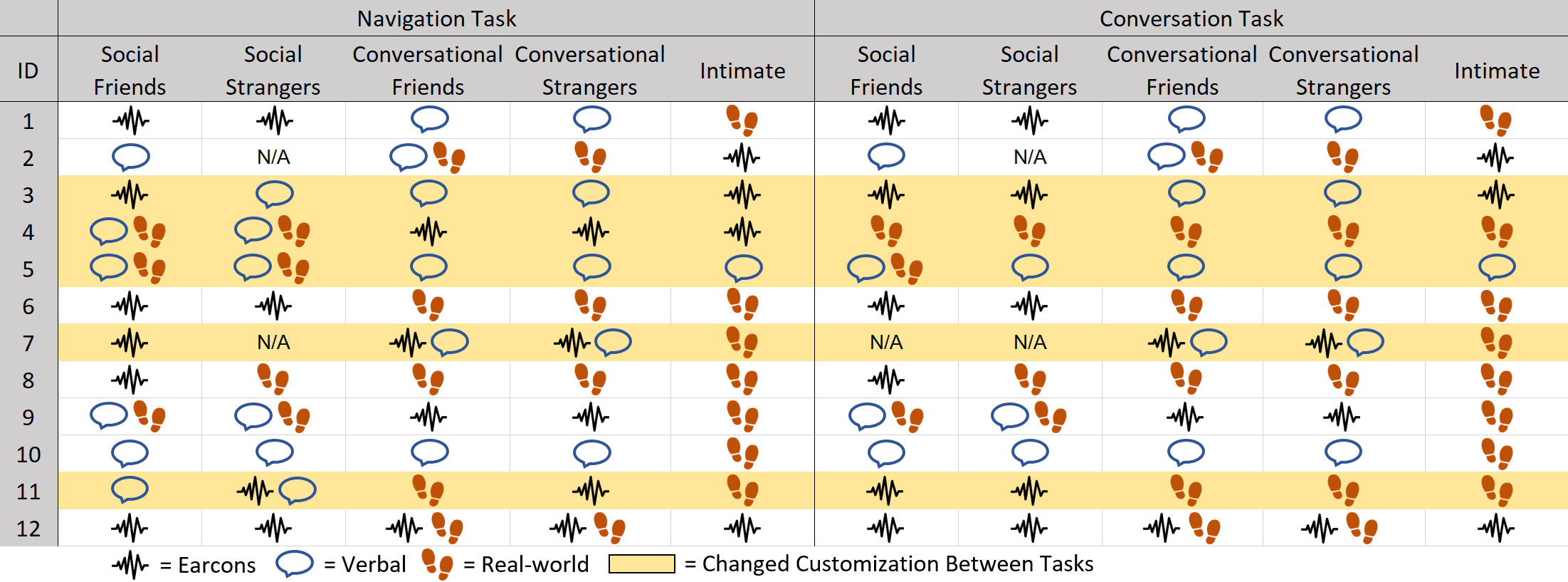}
    \caption{Chart of participants' VRBubble customization in navigation and conversation tasks. }
    \label{fig:customizetable}
    \Description{Listing selected customization options in order of: Participant, Navigation Task, Friends at the social bubble, Strangers at the social bubble, Friends at the conversational bubble, Strangers at the conversational bubble, Intimate bubble. Then Conversation Task and the corresponding customizations if there was a change in customization between tasks.
    E will stand for "earcon", V will stand for "verbal", R will stand for "real-world", and N will stand for "none"
    P1, Navigation Task, E, E, V, V, R.
    P2, Navigation Task, V, N, V + R, R, E.
    P3, Navigation Task, E, V, V, V, E, Conversation Task, E, E, V, V, E.
    P4, Navigation Task, V + R, V + R, E, E, E, Conversation Task, R, R, R, R, R.
    P5, Navigation Task, V + R, V + R, V, V, V, Conversation Task, V + R, V, V, V, V.
    P6, Navigation Task, E, E, R, R, R.
    P7, Navigation Task, E, N, E + V, E + V, R, Conversation Task, N, N, E + V, E + V, R.
    P8, Navigation Task, E, R, R, R, R.
    P9, Navigation Task, V + R, V + R, E, E, R.
    P10, Navigation Task, V, V, V, V, R.
    P11, Navigation Task, V, E + V, R, E, R, Conversation Task, E, E, R, R, R.
    P12, Navigation Task, E, E, E + R, E + R, E.}
\end{figure*}

Participants customized the audio alternatives for different social tasks, avatar roles, and bubbles (Figure \ref{fig:customizetable}).  Figure \ref{fig:navcustomize} shows the distribution of audio alternatives chosen for the navigation task, and Figure \ref{fig:convocustomize} shows the distribution for the conversation task. While no participants selected the exactly same combinations, we saw some patterns across participants that we report below.

\textit{\textbf{Navigation vs. Conversation Task.}}
In general, we found that verbal notification was most popular for the navigation task, while earcons and real-world sound effects were more popular in the conversation task (except for Intimate Bubble), as shown between Figure \ref{fig:navcustomize} and Figure \ref{fig:convocustomize}. While seven participants used the same combination for both navigation and conversation tasks, five participants changed their customization. We found that participants tended to simplify their audio combination in the conversation task, and all changes happened in Social Bubble. The changes included removing an audio alternative (P4, P11 removed verbal notification, P5 removed real-world sound effect for strangers, and P7 removed earcons for friends), and changing verbal notifications to more brief earcons (P3 for strangers, P11 for friends).  
This pattern suggested that users preferred shorter and non-verbal feedback during audio-focused tasks like conversation, whereas more informative feedback was preferred for exploration-based tasks like navigation.



\textit{\textbf{Friends vs Strangers.}}
Participants had divergent preferences when distinguishing friends and strangers. Some participants (P2, P5, P7, P8) chose more detailed information for friends and felt that stranger information was less important especially in the Social Bubble. P2 and P7 even did not want any audio feedback for strangers in the Social Bubble. On the other hand, some participants (P3, P11) wanted to know more about strangers: P3 selected earcons for friends and more informative verbal notification for strangers in the Social Bubble in the navigation task, while P11 added earcons for the strangers to alert himself. They explained that in a real use case, users were likely to already know enough information about friends and would only need more detailed information about strangers. As P3 described, \textit{``The verbal would be useful because that might indicate to me it's a stranger I don't know. And this is their name.''} 



\textit{\textbf{Patterns across Bubbles.}}
Most participants preferred the real-world sound effects for the Intimate Bubble, while some participants selected the abstract earcons. Participants mostly made the decision based on which audio feedback they found more pleasant. Interesting, P5 was the only participant who expressed interest in knowing who they bumped into, thus choosing verbal notification. This may suggest that in social VR, avatar collision was an event that PVI wanted to avoid but not important enough for PVI to care about the details.

For the Social and Conversation Bubbles, we found that participants had different preferences for different bubbles without clear patterns, and only P10 used verbal notification for both bubbles.


\textit{\textbf{Audio Combinations.}}
The most popular audio combination was between verbal notifications and real-world sound effects, which was adopted by four participants (P4, P5, and P9 used the combination for the Social Bubble, P2 used it at the Conversation Bubble). Two participants (P7, P11) also combined earcons with verbal notifications. Participants used such combinations to add immersion to the detailed information provided by verbal notifications. As P5 described, \textit{``Because that will add more reality to what I'm doing, instead of just a boring computer thing.''} 
Only P12 used a combination of earcons and real-world footsteps. As he explained, \textit{``The abstract [earcon] is my favorite and that's why I chose most of the options here. But I thought it would be helpful to have the footsteps as well just so it's a little bit different [from just the earcon].''}

\section{Discussion}

With VRBubble, we contributed the first VR technique that leveraged the social distances (i.e., Bubbles) in Hall's proxemic theory \cite{hall} to enhance PVI's peripheral awareness of avatars in a complex and dynamic social VR context. VRBubble enabled user customization by providing three spatial audio alternatives---earcons, verbal notifications, and real-world sound effects---and allowing users to select and combine their preferred audio feedback for different bubbles, avatars, and social contexts.

Our evaluation with 12 PVI demonstrated the effectiveness of VRBubble: It significantly enhanced participants' awareness of the amount of avatars they passed by in a navigation task and enabled participants to identify most avatars, including their names and relationship to the users in both navigation and conversation tasks.   
However, we found that VRBubble could be distracting especially in crowded environments with high amount of avatars. Our research was the first step towards the general accessibility of the dynamic social VR. Based on our exploration, we discuss the takeaways, design implications, technology generalizability, and the lessons we learnt from our remote study. 

\subsection{Audio Preferences for Peripheral Awareness}
Besides seeing VRBubble as a whole VR technique, our research also explored PVI's experiences and preferences of perceiving peripheral information via different audio modalities. Unlike prior work that studied the use of different audio feedback in various primary tasks (e.g., \cite{morrison2021peoplelens, zhao2018face, anam2014expression}), we focused on using audio to convey peripheral dynamics and investigated the feasibility of different audio modalities in two representative social contexts: navigation and conversation. Our study showed that compared to a fixed design (e.g., nonadjustable audio beacons), flexible audio customization allows users to effectively control distraction from a given primary task. In our study, five participants changed their audio selection after experiencing a different social task (Table 1), highlighting the need for customization depending on the social context. 

Our research identified PVI's customization patterns. First, participants' audio preferences were driven by the cognitive load of the primary task. They preferred more detailed and descriptive audio modality (e.g., verbal notification) in less-attention-demanding tasks, but could only handle brief and distinct audio feedback (e.g., earcons, sound effects) in more attention-demanding and audio-focused tasks. Second, the importance of peripheral information was another factor that influenced PVI's audio selection. Specifically, the more detailed verbal descriptions or multiple audio alternative combinations were selected for more important avatars.  

Our results also suggested (but need further investigation) that people may have a upper limit of capacity to perceive peripheral information in attention-demanding tasks (e.g., perceiving five avatars at most in a conversation task), so that they had to give up on less important peripheral information, such as information on strangers in the Social Bubble. While our research focused on the social VR context, the preferences of audio modalities by PVI to receive peripheral information could be applied to broader use cases, such as situational awareness systems for drivers.

\subsection{Design Implications}
Based on participants' experiences with VRBubble, we distill design implications to inspire more accessible and immersive social VR experience for PVI. 

\textit{\textbf{Context-Adaptive Feedback.}} 
Our study indicated that participants preferred different audio modalities for different social contexts. However, social activity is a highly dynamic process where users' context and primary task can change quickly. For example, a user could be navigating, but stop to chat with a passing friend in the next moment. Thus, it could be difficult for PVI to manually adjust their audio combinations every time to match the fast-changing social contexts. Some participants suggested pre-defining multiple sets of audio combinations for different contexts and designing easy ways to toggle between them. As P9 suggested, \textit{``[You can have] a walking around mode, and a talking conversation mode. If you exit a conversation and you're walking around, switch back to names.''} To reduce the effort of active user control, researchers could consider context-adaptive feedback mechanism, which recognizes users' social context with AI technology and automatically switches to users' preferred feedback based on the current context. 


\textit{\textbf{Balance Automatic and Manually-queried Feedback.}}
VRBubble focused on conveying peripheral information, thus providing proactive feedback that was automatically triggered by environmental changes to reduce users' interaction efforts. However, in our study, several participants indicated the desire for actively querying avatar information since they may have missed some prior information or need more detailed information about particular avatars of interest. We acknowledge that both types of feedback have their values. For example, our study showed that while proactive feedback were suitable for important and urgent information, too much feedback can distract and overwhelm the users. Meanwhile, manually-queried feedback can be a complement to provide detailed information on demand and reduce feedback overload. 
It is thus important to consider how to balance proactive and manually-queried feedback to optimize users' experience in social VR, especially in crowded virtual environments or cognitively heavy tasks. For example, proactive notifications can be used for important and nearby avatars, and manually-queried feedback can serve to provide additional information based on users' interest. Additionally, users should also have the option to manually repeat any information provided by proactive feedback.

\textit{\textbf{Enable Third-Party Audio Design.}}
While providing different audio alternatives, VRBubble provided only pre-defined audio effects, especially for earcons and real-world sound effects. Interestingly, our study indicated that participants had distinct hearing abilities, audio preferences, and audio aesthetic. For example, some participants (P2, P8) found the real-world footstep more pleasant to hear, while some (P4, P9) preferred the more abstract earcons. Even for the same audio modality, participants had different preferences. Some (P3, P6, P11, P12) liked the current earcon design, while some (P1, P4, P5, P9, P10) found them difficult to distinguish. To better fulfill the users' needs, one solution could be allowing users to design and upload their own audio source to represent particular avatars. However, designing and generating audio source could be technically challenging, especially for people with visual impairments. It is thus important to consider the trade-off between the users' effort and satisfaction. Online communities for avatar sonification could also be encouraged to provide PVI a place to search and find their preferred audio source, similar to Thingiverse \cite{thingiverse} which serves as an online community that provides 3D model resources for makers.

\subsection{Generalizability}
We discuss the generalizability of our technique from both the platform and the users' perspectives. While implemented and evaluated on a desktop-based VR platform, VRBubble can be easily transferred to HMD-based VR as it focuses on proactive audio design that is not device or platform specific. It can be adapted to other hardware through combination with suitable input techniques (e.g., input via controllers or body/head movements).

Beyond PVI, VRBubble could also be applied for sighted users in social VR. Various real-world presence cues, such as breathing or wind movement are not conveyed through visual information. Thus providing sighted users with additional peripheral information via audio feedback could enhance immersion in social VR, as well as ensuring users' privacy (i.e., being aware of potential out of sight eavesdroppers).

\subsection{Evaluation via a Remote Study}
We conducted our study remotely due to the restriction of the pandemic. We summarize our experience and the lessons we learnt from the remote study. First, we found it difficult for participants to setup all the software and hardware required by our study independently. To guide participants through the whole setup process, we connected with the participants 30 minutes ahead of the study via Zoom, sent them the url to our VR environment, and asked them to share screen with us during the setup process. Participants generally appreciated our assistance via Zoom and felt the use of Zoom sharing function was effective for the remote study. As P7 noted, \textit{``You've just proven a feature not in virtual reality right now---feature sharing. You just showed how important that is for someone with a disability because accessibility [functionality] doesn't always work.''} 

Second, it is hard to guarantee the consistency of participants' experience in a remote study, depending on each individual's devices and environments. For example, they may receive different audio quality due to the different headphones, and some of them may experience feedback delay due to unstable Internet access.  
For this study, we tried to ensure a certain level of consistency by checking that participants had the pre-requisite equipment, including a keyboard, a pair of headphones that supports spatial audio, and a Microsoft Edge browser installed to access our VR environment. Participants shared their screen and audio via Zoom, so that we can check whether our feature was functioning properly in real time. However, the audio share feature in Zoom did not support spatial audio, which made it impossible to fully confirm users' experience. In the future, how to setup a remote study platform and enable researchers to easily control and confirm users' experience is a vital research direction to explore.  

\subsection{Limitations and Future Work}
Our research has limitations. Since our study focused on the design and evaluation of the audio feedback (i.e., the output) in VRBubble, we used a Wizard-of-Oz \cite{dahlback1993wizardofoz} setup to enable participants to customize the audio alternatives. Thus, how to enable PVI to efficiently and easily control and customize VRBubble's audio modalities (i.e., the input) still mains unaddressed. A future work of interest would be to explore effective input modalities to support fast interaction for PVI in social VR. 

Additionally, our current evaluation used mock-up social VR environments with system-generated avatars, which may not fully reflect the avatar dynamics in the real social VR environments. For example, avatars in social VR may be conversing rather than standing silently, so that they might be located or identified by their voice. This can potentially change users' experience with VRBubble. Future research should evaluate VRBubble in more realistic social VR scenarios and adapt the design to different situations; for example, a context-aware technique that adjusts the audio feedback based on avatars' voice activity (e.g., muting or reducing the volume of audio notifications for a speaking avatar). 

\section{Conclusion}
We presented VRBubble, a social VR technique to augment peripheral awareness by utilizing Hall's proxemic theory. VRBubble provided three audio alternatives---earcons, verbal notification, and real-world sound effects---for PVI to select and combine to maintain aware of different avatars at different social distances. We compared VRBubble to a standard audio beacon baseline via a user study with 12 participants with visual impairments. We assessed the effect of VRBubble on participants' ability to identify avatars as well as its distraction in different primary social tasks. Participants found VRBubble effective at providing previously inaccessible avatar information, and expressed the need for customizing audio feedback in different social VR contexts. Our study contributed a novel accessibility technique to inform the social VR dynamics and provided implications for the future design of accessible VR for people with visual impairments.

\begin{acks}
We thank the National Federation of the Blind for helping us recruit for our study, as well as the anonymous participants that provided their perspective.
\end{acks}


\bibliographystyle{ACM-Reference-Format}
\bibliography{sources.bib}

\appendix

\end{document}